\documentclass[12pt]{article}
\usepackage{amsmath,amssymb,mathrsfs}
\usepackage{graphicx}

\newcommand{\pder}[2]{\frac{\partial #1}{\partial #2}}
\newcommand{\dif}[1]{{d}{#1}}

\newcommand{\Tr}{{\it{\rm Tr\, }}}

\newcommand{\bra}[1]{\left\langle #1\right|}
\newcommand{\ket}[1]{\left| #1\right\rangle}
\def\bb{\langle}
\def\kk{\rangle}

\begin{document}
\begin{titlepage}
\begin{flushright}
KUNS-2398\\
\end{flushright} 
\vspace{15mm}
\begin{center}
{\large \bf
Factorization of the Effective Action in the IIB Matrix Model
}
\end{center}
\vspace{7mm}
\begin{center}
Yuhma A{\sc sano}$^a$\footnote{
yuhma@gauge.scphys.kyoto-u.ac.jp
}
Hikaru K{\sc awai}$^{a}$\footnote{
hkawai@gauge.scphys.kyoto-u.ac.jp
} and
Asato T{\sc suchiya}$^b$\footnote{
satsuch@ipc.shizuoka.ac.jp
}

\par \vspace{7mm}
{\it
$^a$Department of Physics, Kyoto University,\\
Kyoto 606-8502, Japan
}\\
{\it
$^b$Department of Physics, Shizuoka University,\\
836 Ohya, Suruga-ku, Shizuoka 422-8529, Japan
}\\

\end{center}                                                                           
\vspace{7mm}
\begin{abstract}\noindent
We study the low-energy effective action of the IIB matrix model in the derivative interpretation, where
the diffeomorphism invariance is manifest and arbitrary manifolds are described by matrices.
We show that it is expressed as a sum of terms, each of which
is factorized into a product of diffeomorphism 
invariant action functionals:
$S=\sum_{i}c_is_i+\sum_{i,j}c_{ij}s_is_j+\sum_{i,j,k}c_{ijk}s_is_js_k+\cdots .$
Each action functional $s_i$
is an ordinary local action of the form
$s_i=\int \dif{^Dx}\sqrt{-g}\, O_i(x),$
where $O_i(x)$ is a scalar operator.
This is also true for the background consisting of block diagonal matrices.
In this case, the effective action can be interpreted as describing a multiverse where the universes
described by the blocks are connected by wormholes.
\end{abstract}
\setcounter{footnote}{0}
\end{titlepage}


\section{Introduction}
Matrix models have been proposed as candidates for nonperturbative formulation 
of the string theory \cite{BFSS97M-,IKKT97A-,Dijkgraaf:1997vv}.
The IIB matrix model \cite{IKKT97A-} is one of those models. 
While the space-time coordinates are represented by matrices in its original proposal, 
the matrices are interpreted as a covariant derivative
in the alternative interpretation, which we call the derivative interpretation \cite{HKK06De}. 
In other words, the matrices are identified with momenta
and it may be T-dual to the original proposal,
although it is an open problem to realize the T-duality in the matrix model.
The advantage of the derivative interpretation is that the diffeomorphism invariance is manifest and 
arbitrary manifolds are described by matrices, which
enables one to study the structure of the gravity in the model
from an alternative point of view.

More concretely, an arbitrary $D$-dimensional manifold $\mathcal{M}$ with $D\leq 10$ 
is described by the matrices as follows.
First we regard the matrices as linear operators acting on
$C^\infty (E_{prin})$, where $E_{prin}$ is the principal $Spin(D-1,1)$-bundle over $\mathcal{M}$.
The space $C^{\infty}(E_{prin})$ is equivalent to the space of the global section $\Gamma (E_{reg})$.
Here $E_{reg}$ is the vector bundle over $\mathcal{M}$, whose fiber is the space of the regular representation.
Then we identify $D$ matrices out of ten bosonic matrices with
the covariant derivative on $E_{reg}$, and the rest $10-D$ matrices with 0.
The fluctuation around this background includes
infinitely many particles with various spins.
Furthermore, the equations of motion for the background 
agree with the Einstein equation, and the diffeomorphism and
the local Lorentz symmetry are included in the $U(N)$ symmetry of the model.
The interpretation was applied to a supermatrix model which is an extension of the IIB matrix model, where
the local supersymmetry is manifest and the supermatrices are identified with a supercovariant derivative so
that the model is expected to describe supergravity \cite{HKK06Cu}. It was also shown that the interpretation 
allows the model to describe the dilaton and the Kalb-Ramond $B$-field if the torsion
is introduced \cite{FHKK07Fi}.

The quantum effects in the derivative interpretation have not been well understood.
In this paper, we study the low-energy effective action.
We find that the effective action is a sum of terms, each of which
is factorized into a product of diffeomorphism invariant action functionals:
\begin{align}
S=\sum_i c_is_i+\sum_{i,j}c_{ij}s_is_j+\sum_{i,j,k}c_{ijk}s_is_js_k+\cdots .
\label{S_eff}
\end{align}
Here $s_i$ is a space-time integral of a scalar local operator $O_i(x)$:
\begin{align}
s_i=\int \dif{^Dx}\sqrt{-g(x)}\, O_i(x).
\end{align}
This is also true for the background consisting of block diagonal matrices.
The manifold described by such a background consists of some number of universes, each of which
is described by a diagonal block.
It was shown by Coleman over twenty years ago that, in the Euclidean quantum gravity,
the effective action on a multiverse takes this form due to the effect
of wormholes that connect the universes \cite{C88Wh}\footnote{
Recently this type of action was analyzed in the context of the Lorentzian quantum gravity \cite{KO11As,KO11So}.}.
Thus the above effective action of the IIB matrix model is interpreted as describing the multiverse.
The integral over the fluctuation of the matrices corresponds to the effects of
the wormholes.

This paper is organized as follows.
In section 2, we review the derivative interpretation of the IIB matrix model. 
In section 3, we develop 
the method of calculating the effective action and show that it is indeed factorized.
Section 4 is the conclusion and discussion.
Some explicit examples of the calculation are given in the appendices.

\section{Derivative interpretation of the IIB matrix model}
In this section, we review the derivative interpretation of the IIB matrix model 
in which the matrices are interpreted as the covariant derivative \cite{HKK06De}.

The IIB matrix model can be obtained 
by dimensionally reducing ten-dimensional $\mathcal{N}=1\; U(N)$ super Yang-Mills theory to a point.
The action is
\begin{align}
S=\frac{1}{g^2}{\rm Tr}\left[ \frac{1}{4}[A_a ,A_b ][A^a ,A^b ]+\frac{1}{2}\bar \Psi \Gamma ^{a}[A_a ,\Psi ]\right] ,
\label{originalS}
\end{align}
where $a$ and $b$ run from 0 to 9.
$A_a$ is a ten-dimensional Lorentz vector while $\Psi$ is a ten-dimensional Majorana-Weyl spinor.
$A_a$ and $\Psi$ are $N\times N$ hermitian matrices.
This model has manifest $SO(9,1)$ symmetry and the $U(N)$ symmetry.

Arbitrary manifolds with $D \leq 10$ are described by the matrices as follows.
We consider a $D$-dimensional Riemann manifold $\mathcal{M}$
and denote the principal $Spin(D-1,1)$-bundle over it
by $E_{prin}$.
In this paper, we assume that $\mathcal{M}$ has the Lorentzian signature, where 
the metric is given by $\eta _{ab}=\mathrm{diag}(-1,+1,\cdots ,+1)$.
In what follows, we put $G=Spin(D-1,1)$ and denote elements of $\mathcal{M}$ and $G$ by $x$ and $g$, respectively.
Let $\mathcal{M}$ have the open covering $\mathcal{M}=\cup_i U_i$ and the coordinates on $U_i$ be $x_{[i]}$.
Then, $E_{prin}$ is constructed from the set of $U_i\times G$ by identifying $(x_{[i]},g_{[i]})$ and $(x_{[j]},g_{[j]})$
on the overlap region $U_i \cap U_j$, when they satisfy $x_{[i]}=x_{[j]}$ and $g_{[i]}=t_{ij}(x)g_{[j]}$
where $t_{ij} \in G$ is the transition function.
The covariant derivative on $\mathcal{M}$ acts on a function $f \in C^{\infty}(E_{prin})$ as
\begin{align}
\nabla _af^{[i]}(x_{[i]},g_{[i]})=e_a^{[i]\, \mu}(x_{[i]})\left( \partial _\mu -\frac{i}{2}\omega _{\mu}^{[i]\; bc}(x_{[i]})O_{bc}\right) f^{[i]}(x_{[i]},g_{[i]}) ,
\end{align}
where $a,\:b,\:c$ take $0,1,\cdots D-1$.
$e_a^{\;\; \mu}(x)$ is the vielbein, and $\omega _{\mu}^{\;\; bc}(x)$ is the spin connection. 
The suffix $[i]$ indicates the objects on the patch $U_i$.
Note that the function $f$ is globally defined, namely
$f^{[i]}(x_{[i]},g_{[i]})=f^{[j]}(x_{[j]},g_{[j]})$ holds on $U_i\cap U_j$. 
$O_{bc}$ is the $Spin(D-1,1)$ generator satisfying
\begin{align}
(1+i\varepsilon ^{bc}O_{bc})f^{[i]}(x_{[i]},g_{[i]})= f^{[i]}(x_{[i]},(1-i\varepsilon ^{bc}T_{bc})g_{[i]}),
\end{align}
where $T_{bc}$ is the matrix of the generator for the fundamental representation.

Then we define the endomorphic covariant derivative $\nabla _{(a)}$ as
\begin{align}
\nabla _{(a)}^{[i]}:=R_{(a)}^{\;\;\;\;\; b}(g_{[i]}^{-1})\nabla _b^{[i]},
\end{align}
where $R_{(a)}^{\;\;\;\;\; b}(g)$ is the vector representation matrix of $g\in G$. 
$R_{(a)}^{\;\;\;\;\; b}(g^{-1})$ can be viewed as the Clebsch-Gordan coefficients
for 
\begin{align}
V_{vec}\otimes V_{reg}\cong V_{reg}\oplus \cdots \oplus V_{reg},
\end{align}
where $V_{vec}$ and $V_{reg}$ are the vector and the regular representations of $G$, respectively.
Indeed, one can verify that each component of $\nabla _{(a)}$ is an endomorphism on $C^\infty (E_{prin})$ as
\begin{align}
\nabla _{(a)}^{[i]}f(x_{[i]},g_{[i]})&=R_{(a)}^{\;\;\;\;\; b}(g_{[i]}^{-1})\nabla _b^{[i]}f(x_{[i]},g_{[i]})\nonumber \\
&=R_{(a)}^{\;\;\;\;\; b}((t_{ij}g_{[j]})^{-1})R_{b}^{\;\;\; c}(t_{ij})\nabla _c^{[j]}f(x_{[j]},g_{[j]})\nonumber \\
&=R_{(a)}^{\;\;\;\;\; b}(g_{[j]}^{-1})\nabla _b^{[j]}f(x_{[j]},g_{[j]})\nonumber \\
&=\nabla _{(a)}^{[j]}f(x_{[j]},g_{[j]}).
\end{align}
Note that the index $(a)$ in $\nabla _{(a)}$ is only a label which has nothing to do with
the local Lorentz transformation. 
More generally, multiplying an index $\beta$ of the representation $r$
by the Clebsch-Gordan coefficients $R_{\bb r\kk \beta}^{\;\;\;\;\;\; (\alpha )}(g)$ yields a non-transforming index $(\alpha )$.
For instance, for a field of the representation $r$, $f_\beta (x)$,
an element $h\in G$ acts as
\begin{align}
R_{\bb r\kk (\alpha)}^{\;\;\;\;\;\;\;\;\; \beta}(g^{-1})f_\beta (x)\longmapsto &\, R_{\bb r\kk (\alpha )}^{\;\;\;\;\;\;\;\;\; \beta}(g^{-1})R_{\beta}^{\;\;\; \gamma}(h)f_\gamma (x)\nonumber \\
=&\, R_{\bb r\kk (\alpha )}^{\;\;\;\;\;\;\;\;\; \beta}((h^{-1}g)^{-1})f_\beta (x).
\end{align}
Thus, $R_{\bb r\kk (\alpha )}^{\;\;\;\;\;\;\;\;\; \beta}(g^{-1})f_\beta (x)$ behaves as the regular representation
for each $(\alpha )$.
The indices with parentheses $(\;)$ are mere labels that are invariant under 
the local Lorentz transformation.
We convert the indices with and without parentheses by $R_{\bb r\kk (\alpha )}^{\;\;\;\;\;\;\;\;\; \beta}(g^{-1})$.

Here we interpret the matrices as endomorphisms on $C^\infty (E_{prin})$.
Thus we can construct a matrix configuration which describes the manifold $\mathcal{M}$.
More explicitly, we identify
$D$ matrices out of the ten $A_a$'s
with $\nabla _{(a)}$, and the other $10-D$ matrices with 0:
\begin{align}
A_a=\left\{
\begin{array}{ll}
i\nabla _{(a)}&(a=0,\cdots ,D-1)\\
0&(a=D,\cdots ,9)
\end{array}
\right. .
\label{ansatz}
\end{align}
In general, $\mathcal{M}$ can consist of some number of connected components, each of which describes a universe.
Such a manifold is described by block diagonal matrices.

The Einstein equation follows from the equations of motion of the matrix model.
We take the ansatz \eqref{ansatz} with the fermion vanishing.
Then, the equations of motion
\begin{align}
[A^b,[A_a,A_b]]=0
\end{align}
become
\begin{align}
-i[\nabla ^{(b)},[\nabla _{(a)},\nabla _{(b)}]]=0.
\label{EoM}
\end{align}
Using the formula $[\nabla _{a},\nabla _{b}]=-\frac{i}{2}R_{ab}^{\;\;\;\; cd}O_{cd}$, we find
\begin{align}
-i[\nabla ^{(b)},[\nabla _{(a)},\nabla _{(b)}]]
=-\frac{1}{2}R^{\;\;\;\;\,  a^\prime}_{(a)}(g^{-1})\left\{ (\nabla ^{b}R_{a^\prime b}^{\;\;\;\;\; cd})O_{cd}+2iR_{a^\prime}^{\;\;\;\, d}\nabla _{d}\right\} ,
\end{align}
which result in
\begin{align}
R_{ab}=0,\;\; \nabla ^{b}R_{ab}^{\;\;\;\; cd}=0.
\end{align}
Because the second equation follows from the first one, we find that \eqref{EoM} is equivalent to the Einstein equation in the vacuum.

In general, the matrices $A_a$ and $\Psi$ can be expanded 
in terms of the differential operators $\nabla _{(a)}$ and $O_{(b)(c)}$.
The expansion of $A_a$ takes the form\footnote{
Later, for simplicity of notation, we make a replacement 
$A_{(a)}^{\{ 0\}}\to B_{(a)}$, $A_{(a)}^{\{ 0\} bc}\to \varpi _{(a)}^{\;\;\;\; bc}$, and $A_{(a)}^{\{ 1\} b}\to h_{(a)}^{\;\;\;\; b}$.
}
\begin{align}
A_{(a)}(x,g;y,h)&=\Bigg[ i\nabla _{(a)}+A_{(a)}^{\{ 0\}}(x,g)+\frac{1}{4}\{ A_{(a)}^{\{ 0\} bc}(x,g),O_{bc}\} \nonumber \\
&\;\;\;\;\;\;\;\;\;\;\;\;\; +\frac{1}{16}\{ A_{(a)}^{\{ 0\} bcde}\! (x,g),O_{bc}O_{de}\} +\cdots \nonumber \\
&\;\;\;\; +\frac{1}{2}\Big\{ A_{(a)}^{\{ 1\} b}(x,g)+\frac{1}{4}\{ A_{(a)}^{\{ 1\} bcd}\! (x,g),O_{cd}\} +\cdots ,i\nabla _{b}\Big\} \nonumber \\
&\;\;\;\;\; +\frac{1}{4}\Big\{ A_{(a)}^{\{ 2\} bc}\! (x,g)+\frac{1}{4}\{ A_{(a)}^{\{ 2\} bcde}\! (x,g),O_{de}\} +\cdots ,-\nabla _{b}\nabla _{c}\Big\} \nonumber \\
&\;\;\;\;\;\;\;\;\;\;\;\;\;\;\;\;\;\;\;\;\;\;\;\;\;\;\;\;\;\;\;\;\;\;\;\;\;\;\;\;\;\;\;\;\;\;\;\;\;\;\;\;\;\;\;\;\;\;\;\;\;\;\;\;\;\;\;\;\;\;\;\;\;\;\; +\cdots \cdots \Bigg]\delta (x-y)\delta _{gh},
\label{Aexp}
\end{align}
where the anti-commutator $\{ \; ,\, \}$ is introduced to make $A_a$ hermitian.
The leading term is the classical solution, which corresponds to the Ricci-flat covariant derivative 
as we have seen above. 
By using the Peter-Weyl theorem, we further expand the fields appearing in the subleading terms and
obtain the fields with various spins.
For instance,
\begin{align}
A_{(a)}^{\{ 0\}}(x,g)&=R^{\;\;\;\;\,  b}_{(a)}(g^{-1})A_{a}^{\{ 0\}}(x,g)\nonumber \\
&=R^{\;\;\;\;\,  b}_{(a)}(g^{-1})\sum_{r:\mathrm{irr.}}R_{\bb r\kk \beta}^{\;\;\;\;\;\; (\alpha )}\! (g)\, \tilde A_{\langle r\rangle \, b\, (\alpha )}^{\{ 0\}\, \beta}(x),
\label{PW}
\end{align}
where $r$ is a irreducible representation and $\alpha ,\beta$ are its indices.
$\sum_{r:\text{irr.}}$ represents the summation over all irreducible representations.
In section 3, we abbreviate $\tilde A_{\bb r\kk \, b\, (\alpha )}^{\{ 0\}\, \beta}(x)$ as $A_{(I)J}(x)$, where $(I)$ and $J$ stand for sets of indices with and without parentheses, respectively.

The model is invariant under the unitary transformation $U(N)$.
As we will see below, it
includes the $U(1)$ gauge transformation, the local Lorentz transformation, 
and the diffeomorphism.
The infinitesimal $U(N)$ transformation is expressed as
\begin{align}
\delta A_a&=i[\Lambda ,A_a],
\label{U(N) transformation for A_a}\\
\delta \Psi &=i[\Lambda ,\Psi ],
\end{align}
by using an infinitesimal matrix $\Lambda$,
which can be expanded as follows:
\begin{align}
\Lambda (x,g;y,h)&=\Bigg[ \lambda ^{\{ 0\}}(x,g)+\frac{1}{4}\{ \lambda ^{\{ 0\} bc}(x,g),O_{bc}\} \nonumber \\
&\;\;\;\;\;\;\;\;\;\;\;\;\; +\frac{1}{16}\{ \lambda ^{\{ 0\} bcde}\! (x,g),O_{bc}O_{de}\} +\cdots \nonumber \\
&\;\;\;\; +\frac{1}{2}\Big\{ \lambda ^{\{ 1\} b}(x,g)+\frac{1}{4}\{ \lambda ^{\{ 1\} bcd}\! (x,g),O_{cd}\} +\cdots ,i\nabla _{b}\Big\} \nonumber \\
&\;\;\;\;\; +\frac{1}{4}\Big\{ \lambda ^{\{ 2\} bc}\! (x,g)+\frac{1}{4}\{ \lambda ^{\{ 2\} bcde}\! (x,g),O_{de}\} +\cdots ,-\nabla _{b}\nabla _{c}\Big\} \nonumber \\
&\;\;\;\;\;\;\;\;\;\;\;\;\;\;\;\;\;\;\;\;\;\;\;\;\;\;\;\;\;\;\;\;\;\;\;\;\;\;\;\;\;\;\;\;\;\;\;\;\;\;\;\;\;\;\;\;\;\;\;\;\;\;\;\;\;\;\;\;\;\;\;\;\;\;\; +\cdots \cdots \Bigg]\delta (x-y)\delta _{gh}.
\end{align}

We further expand $\lambda^{(0)}(x,g)$, $\lambda^{\{0\} bc}(x,g)$ and $\lambda^{\{1\} b}(x,g)$,
using the Peter-Weyl theorem, and denote
the coefficients for the trivial representations by 
$\lambda^{(0)}(x)$, $\lambda^{\{0\} bc}(x)$ and $\lambda^{\{1\} b}(x)$, respectively.
They correspond to
the $U(1)$ gauge transformation, the local Lorentz transformation and the diffeomorphism, respectively.
Indeed,
for $\lambda ^{\{ 0\}}$, we have
\begin{align}
i[\lambda ^{\{ 0\}}(x),i\nabla _{(a)}]=\nabla _{(a)}\lambda ^{\{ 0\}}(x),
\end{align}
and \eqref{U(N) transformation for A_a} implies 
\begin{align}
\delta \tilde A_{a}^{\{ 0\}}(x)=\nabla _{a}\lambda ^{\{ 0\}}(x)+\text{(higher spin fields)},
\label{dAU(1)}
\end{align}
where $\tilde A_{a}^{\{ 0\}}(x)$ is the coefficient for the trivial representation in (\ref{PW}).
The second term in the right-hand side stands for the effect of the higher spin fields.
If they are massive, they do not contribute in the low energy.
Thus, \eqref{dAU(1)} is the $U(1)$ transformation for $\tilde A_{a}^{\{ 0\}}(x)$.
We will ignore such higher spin terms in the following.
For $\lambda ^{\{ 0\} bc}$, we obtain from $i[\{ \lambda ^{\{ 0\} bc}(x),O_{bc}\} ,i\nabla _{(a)}]$ that
\begin{align}
\delta e_{a}^{\; \mu}(x)&=\lambda _{\;\;\;\, a}^{\{ 0\} \; b}(x)\, e_b^{\; \mu}(x),\\
\delta \omega_\mu ^{\;\; bc}(x)&=e^a_{\; \mu}\nabla _{a}\lambda ^{\{ 0\} bc} (x).
\end{align}
These are the local Lorentz transformation for $e_{a}^{\; \mu}(x)$ and $\omega_\mu ^{\;\; bc}(x)$.
The transformation of $\tilde A_{c}^{\{ 0\}}(x)$ is given by
\begin{align}
\delta \tilde A_{c}^{\{ 0\}}(x)=\frac{i}{4}[\{ \lambda ^{\{ 0\} ab}(x),O_{ab}\} ,\tilde A_{c}^{\{ 0\}}(x)]=\lambda _{\;\;\;\, c}^{\{ 0\} \; b}(x) \tilde A_{b}^{\{ 0\}}(x),
\end{align}
which is indeed the local Lorentz transformation for $\tilde A_{c}^{\{ 0\}}(x)$.
For $\lambda ^{\{ 1\} b}$, we obtain from \\
$i[\{ \lambda ^{\{ 1\} b}(x),i\nabla _{b}\} ,i\nabla _{(a)}]$ that
\begin{align}
\delta e_{a}^{\; \mu}(x)&=(\nabla _a\lambda ^{\{ 1\} b}(x))\, e_b^{\; \mu}(x), \label{diff for e}\\
\delta \omega_\mu ^{\;\; bc}(x)&=-\lambda ^{\{ 1\} \nu}(x)\, R_{\nu \mu}^{\;\;\;\; bc}(x).
\end{align}
These are the diffeomorphism for $e_{a}^{\; \mu}(x)$ and $\omega_\mu ^{\;\; bc}(x)$,
where $x^\mu \to x'^\mu =x^\mu +\lambda ^{\{ 1\} \mu}$.
We also obtain
\begin{align}
\delta \tilde A_a^{\{ 0\}}(x)=\frac{i}{2}[\{ \lambda ^{\{ 1\} b}(x),i\nabla _{b}\} ,\tilde A_{a}^{\{ 0\}}(x)]=-\lambda ^{\{ 1\} \nu}(x)\nabla _\nu \, \tilde A_a^{\{ 0\}}(x),
\label{diffeo}
\end{align}
which implies together with (\ref{diff for e}) that
\begin{align}
\delta \tilde A_\mu ^{\{ 0\}}(x)=-\lambda ^{\{ 1\} \nu}(x)\nabla _\nu \, \tilde A_\mu ^{\{ 0\}}(x)-(\nabla _\mu \lambda ^{\{ 1\} \nu}(x))\tilde A_\nu ^{\{ 0\}}(x).
\end{align}
This is the diffeomorphism for $ \tilde A_\mu ^{\{ 0\}}(x)$.

The matrices can be viewed as bi-local fields on $E_{prin}$.
Let $\ket{x,g}$ be a basis in $C^{\infty}(E_{prin})$ that satisfies the completeness relation
\begin{align}
1=\int \ket{x,g}\, \dif{^Dx}\, \dif{g}\bra{x,g} ,
\end{align}
where $\dif{g}$ is the Haar measure.
Then, the elements of the matrices $A_a$ are expressed as $A_{(a)}(x,g;y,h)=\bra{x,g}A_a\ket{y,h}$, which
are indeed bi-local fields on $E_{prin}$.

\section{Effective action and its factorization}
In this section, we develop the background field method for the derivative interpretation of the IIB matrix model,
and show that the effective action is a sum of terms, each of which is
a product of local actions.

\subsection{Background field method}
Let us decompose the matrices as
\begin{align}
A_a(x,g;y,h)&=A_{(a)}^{0}(x,g;y,h)+\phi _{(a)}(x,g;y,h),
\label{Afexp}
\end{align}
where $A_{(a)}^0$ is the background and $\phi_{(a)}$ is the fluctuation which will be integrated out.
We further expand the background around the flat metric $e_a^{\; \mu}(x)=\delta _a^\mu$:
\begin{align}
A_{(a)}^{0}(x,g;y,h)&=\Bigg[ i\partial _{(a)}+B_{(a)}(x,g)+\frac{1}{2}\{ h_{(a)}^{\;\;\;\; b}(x,g),i\partial _{b}\} \nonumber \\
&\;\;\;\;\;\;\;\;\;\;\;\;\;\;\;\;\;\;\;\;\;\;\; +\frac{1}{4}\{ \varpi _{(a)}^{\;\;\;\; bc}(x,g),O_{bc}\} +\cdots \Bigg]\delta (x-y)\delta _{gh},
\label{A0exp}
\end{align}
where $B_{(a)},\; h_{(a)}^{\;\;\;\; b},\; \varpi _{(a)}^{\;\;\;\; bc},\cdots$ are functions of $x$ and $g$. We call them background fields.
$\delta _{gh}$ is the delta function on the group manifold, which is defined by 
\begin{align}
\int \dif{h}\delta _{gh}f(h)=f(g).
\end{align}
We do not expand $\phi _{(a)}$ as $A_{(a)}^{0}$, but treat it as a bi-local field.
We also decompose the fermionic field $\Psi$ in a similar manner.

Substituting \eqref{Afexp} into the action, we obtain
\begin{align}
S&=\frac{1}{4}\Tr \Big[ [A_{(a)}^{0},A_{(b)}^{0}]^2+4[A^{0(a)},A^{0(b)}][A_{(a)}^{0},\phi _{(b)}]\nonumber \\
&\;\;\;\;\;\;\; +2[A_{(a)}^{0},\phi _{(b)}]^2+2[A^{0(a)},A^{0(b)}][\phi _{(a)},\phi _{(b)}]-2[A_{(a)}^{0},\phi _{(b)}][A^{0(b)},\phi ^{(a)}]\nonumber \\
&\;\;\;\;\;\;\;\; +4[A^{0(a)},\phi ^{(b)}][\phi _{(a)},\phi _{(b)}]+[\phi _{(a)},\phi _{(b)}]^2
+\text{fermion} \Big] ,
\label{Sexp}
\end{align}
where the irrelevant coupling constant $g$ is set to $1$.
The first term corresponds to the classical part, and the second term is to be dropped in the background field method.

\begin{figure}[htbp]
\begin{center}
\includegraphics[scale=0.77]{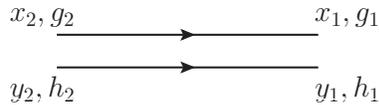}
\caption{\small A propagator between bi-local positions $(x_1,g_1;y_1,h_1)$ and $(x_2,g_2;y_2,h_2)$. 
The two lines live either in different universes or in the same universe.}
\label{fig:pg}
\end{center}
\end{figure}

We would like to know the general form of the effective action, 
which can be obtained by integrating the fluctuations.
As we will see, the mechanism of the factorization does not depend on the spin or statistics of the matrices.
Therefore it is enough to consider a scalar matrix $\phi$ whose quadratic part is given by
\begin{align}
S_{\phi ^2}=\frac{1}{2}\, \Tr \left[ [A^{0(a)},\phi ][A^0_{(a)},\phi ]\right] .
\label{Sphi}
\end{align}
For the one-loop calculation, we keep only the quadratic terms. 
For convenience, we introduce a notation
\begin{align}
\xi ^{(a)}&:=R^{(a)}_{\;\;\;\;\; b}(g^{-1})x^b,
\label{xi}\\
\eta ^{(a)}&:=R^{(a)}_{\;\;\;\;\; b}(h^{-1})y^b.
\label{eta}
\end{align}
Then we can calculate the factor $[A_{(a)}^{0},\phi ]$ in (\ref{Sphi}) as follows:
\begin{align}
\bra{x,g}[A^0_{(a)},\phi ]\ket{y,h}&=\Big( B_{(a)}(x,g)-B_{(a)}(y,h)\nonumber \\
&\;\;\;\;\; +\frac{1}{2}\{ R_{(a)}^{\;\;\;\; b}(g^{-1})+h_{(a)}^{\;\;\;\; b}(x,g),i\pder{}{x ^{b}}\} \nonumber \\
&\;\;\;\;\;\;\;\;\;\;\;\;\;\; +\frac{1}{2}\{ R_{(a)}^{\;\;\;\; b}(h^{-1})+h_{(a)}^{\;\;\;\; b}(y,h),i\pder{}{y ^{b}}\}\nonumber \\
&\;\;\;\;\;\; +\frac{1}{4}\{ \varpi _{(a)}^{\;\;\;\; bc}(x,g),O_{bc}^{(g)}\} +\frac{1}{4}\{ \varpi _{(a)}^{\;\;\;\; bc}(y,h),O_{bc}^{(h)}\} +\cdots  \Big) \, \phi (x,g;y,h)\nonumber \\
&=\left( i\pder{}{\xi ^{(a)}}+i\pder{}{\eta ^{(a)}}+\mathcal{A}_{(a)}(x,g;y,h)\right) \phi (x,g;y,h),
\end{align}
where
\begin{align}
\mathcal{A}_{(a)}(x,g;y,h):&=B_{(a)}(x,g)-B_{(a)}(y,h)\nonumber \\
&\;\;\;\;\; +\frac{1}{2}\{ h_{(a)}^{\;\;\;\; b}(x,g),i\pder{}{x ^{b}}\} +\frac{1}{2}\{ h_{(a)}^{\;\;\;\; b}(y,h),i\pder{}{y ^{b}}\}\nonumber \\
&\;\;\;\;\;\; +\frac{1}{4}\{ \varpi _{(a)}^{\;\;\;\; bc}(x,g),O_{bc}^{(g)}\} +\frac{1}{4}\{ \varpi _{(a)}^{\;\;\;\; bc}(y,h),O_{bc}^{(h)}\} +\cdots .
\label{calA}
\end{align}
Here $O_{bc}^{(g)}$ and $O_{bc}^{(h)}$ act on $g$ and $h$, respectively.
Using the hermiticity of  $\phi$, $\phi (y,h;x,g)=\phi ^\ast (x,g;y,h)$, 
(\ref{Sphi}) is rewritten as 
\begin{align}
S_{\phi ^2}&=-\frac{1}{2}\int \dif{^Dx}\, \dif{^Dy}\, \dif{g}\, \dif{h}\bigg[ \left( \pder{}{\xi _{(a)}}+\pder{}{\eta _{(a)}}-i\mathcal{A}^{(a)}(y,h;x,g)\right) \phi ^\ast (x,g;y,h)\nonumber \\
&\;\;\;\;\;\;\;\;\;\;\;\;\;\;\;\;\;\;\;\;\;\;\;\;\;\;\;\;\;\;\;\;\;\;\;\; \times \left( \pder{}{\xi ^{(a)}}+\pder{}{\eta ^{(a)}}-i\mathcal{A}_{(a)}(x,g;y,h)\right) \phi (x,g;y,h)\bigg] .
\label{Sphi2}
\end{align}
We can read off the Feynman rule from (\ref{Sphi2})\footnote{
In this paper, we use the position-space representation and the double line notation.}. 
It is convenient to express the free propagator (see Figure \ref{fig:pg}) in terms of $\xi$ and the relative coordinate $\xi-\eta$ \cite{Kawai:2009vb,Kawai:2010sf}.
Then, the propagator takes the form
\begin{align}
\mathcal{G}(x_1,g_1;y_1,h_1|x_2,g_2;y_2,h_2) 
&\equiv \bb \phi (x_1,g_1;y_1,h_1)\phi ^\ast (x_2,g_2;y_2,h_2)\kk \nonumber \\
&=G(\xi _1-\xi _2)\delta ((\xi _1-\eta _1)-(\xi _2 -\eta _2 ))\delta _{g_1g_2}\delta _{h_1h_2},
\label{propa}
\end{align}
where $G(\xi _1-\xi _2)$ is the ordinary propagator for a massless scalar field.
Substituting (\ref{xi}) and (\ref{eta}) into (\ref{propa}),
we can further rewrite (\ref{propa}) as
\begin{align}
&\mathcal{G}(x_1,g_1;y_1,h_1|x_2,g_2;y_2,h_2)\nonumber \\
&=G(x_1-x_2)
\delta(R^{(a)}_{\;\;\;\;\;b}(g_1^{-1})(x_1^b-x_2^b)-R^{(a)}_{\;\;\;\;\;b}(h_1^{-1})(y_1^b-y_2^b))\delta_{g_1g_2}\delta_{h_1h_2},
\label{propa2}
\end{align}
where we have used the facts that (\ref{propa}) includes $\delta_{g_1g_2}\delta_{h_1h_2}$ and that 
$G$ is Lorentz invariant.

\subsection{Factorization}
We consider the background which represents a $D$-dimensional manifold.
Here the background may consist of some number of diagonal blocks,
each of which describes a universe.
We show that the effective action is given by
\begin{align*}
S=\sum_i c_is_i+\sum_{i,j}c_{ij}s_is_j+\sum_{i,j,k}c_{ijk}s_is_js_k+\cdots .
\end{align*}
A typical example of the quadratic terms is
\begin{align}
\int \dif{^Dx }\, \sqrt{-g(x)} R(x)\; \int \dif{^Dy }\, \sqrt{-g(y)},
\label{rootgRrootg}
\end{align}
which can arise from the interaction between two blocks
or the self-interaction of one block.
If the two factors are defined on different blocks, this form is naturally expected.
This is because the effective action is invariant under the block diagonal $U(N)$ transformation, each block of which includes the diffeomorphism 
on the corresponding universe.
We will show that the form like \eqref{rootgRrootg} holds even if $x$ and $y$ lie in the same universe.
We use the background field method to show it.
Some examples of explicit calculations are given in the appendices.

First, the classical contribution to the effective action which is the first term in \eqref{Sexp}
has the form
\begin{align}
\Tr [A_{(a)}^{0},A_{(b)}^{0}]^2=\int \dif{^Dx}\, \dif{g}\, [A_{(a)}^{0},A_{(b)}^{0}]^2\! (x,g;x,g).
\end{align}
Although this is naively divergent, it can be evaluated by using the heat-kernel regularization \cite{H06Re}.
The result is a sum of local actions $\sum_i c_is_i$.
This should be invariant under the diffeomorphism and gauge transformation, because they are parts of the original $U(N)$ symmetry.

\begin{figure}[htbp]
\begin{center}
\includegraphics[scale=0.6]{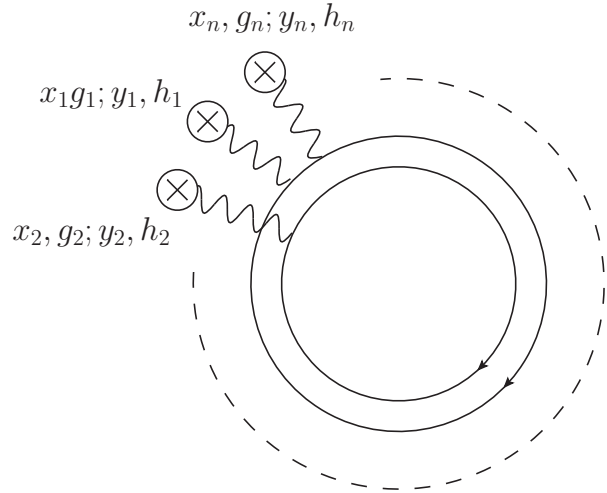}
\caption{\small The general one-loop diagram with $n$ insertions.}
\label{fig:n-vertex}
\end{center}
\end{figure}

\begin{figure}[htbp]
\begin{center}
\includegraphics[scale=0.7]{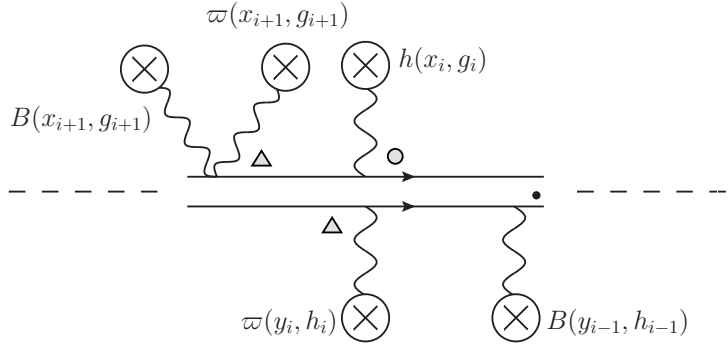}
\caption{\small An example of a part of a diagram. $\triangle$, {\Large $\circ$}, and $\bullet$ indicate the sides on which $O_{bc}$, the derivative $\partial _{b}$ accompanying the background field $h_{a}^{\;\; b}$,
and the derivative $(\pder{}{\xi}+\pder{}{\eta})$ in the kinetic part act, respectively.}
\label{fig:vertexeg}
\end{center}
\end{figure}

Next, we examine the one-loop contribution to the effective action.
The general one-loop diagram with $n$ insertions are shown in Figure \ref{fig:n-vertex}, 
where the two index loops are specified by $(x,g)$ and $(y,h)$, respectively. 
The vertices yield insertions of the background fields.
We can read off the effect of the insertions from \eqref{calA} and \eqref{Sphi2}.
An example is depicted in Figure \ref{fig:vertexeg},
which has a factor of
\begin{align}
&\left[ iB_{(a)}(y_{i-1},h_{i-1})\frac{-i}{2}\{ h^{(a')d'}(x_i,g_i),i\pder{}{x _i^{d'}}\} \mathcal{G}(x_{i-1},g_{i-1};y_{i-1},h_{i-1}|x_{i},g_{i};y_{i},h_{i})\right] \nonumber \\
&\;\; \times \bigg[ \frac{-i}{4}\{ \varpi _{(a')}^{\;\;\;\; b'c'}(y_i,h_i),O_{b'c'}^{(h_i)}\} \nonumber \\
&\;\;\;\;\;\;\;\;\;\;\; \frac{-i}{4}\{ \varpi ^{(a'')b''c''}(x_{i+1},g_{i+1}),O_{b''c''}^{(g_{i+1})}\} \mathcal{G}(x_{i},g_{i};y_{i},h_{i}|x_{i+1},g_{i+1};y_{i+1},h_{i+1})\bigg] .
\end{align}
In general, each propagator is multiplied by some of the following functions or derivatives:
\begin{align}
&A_{(I)J}(x_1),\cdots ,A_{(I)J}(x_n), 
&&\pder{}{x_1^a},\cdots ,\pder{}{x_n^a},\nonumber \\
&R_{\bb r\kk \beta}^{\;\;\;\;\;\; (\alpha )}\! (g_1),\cdots R_{\bb r\kk \beta}^{\;\;\;\;\;\; (\alpha )}\! (g_n), &&O_{ab}^{(g_1)},\cdots ,O_{ab}^{(g_n)},
\label{fn der x}\\
&A_{(I)J}(y_1),\cdots ,A_{(I)J}(y_n), 
&&\pder{}{y_1^a},\cdots ,\pder{}{y_n^a},\nonumber \\
&R_{\bb r\kk \beta}^{\;\;\;\;\;\; (\alpha )}\! (h_1),\cdots R_{\bb r\kk \beta}^{\;\;\;\;\;\; (\alpha )}\! (h_n), 
&&O_{ab}^{(h_1)},\cdots ,O_{ab}^{(h_n)}.
\label{fn der y}
\end{align}
where $A_{(I)J}$'s are the background fields given by \eqref{PW}.
Then the general $n$-vertex loop can be expressed as a sum of terms of the form
\begin{align}
\int \dif{^Dx_1}\cdots \dif{^Dx_n}\, \dif{^Dy_1}\cdots \dif{^Dy_n}\, \dif{g_1}\cdots \dif{g_n}\, \dif{h_1}\cdots \dif{h_n}\, \prod_{i=1}^{n}\mathcal{P}_i.
\label{n-vertex loop}
\end{align}
$\mathcal{P}_i$ is given by
\begin{align}
\mathcal{P}_i&=\mathcal{F}_i\, \mathcal{F}'_i\, \mathcal{G}(x_{i},g_{i};y_{i},h_{i}|x_{i+1},g_{i+1};y_{i+1},h_{i+1}),
\end{align}
where the suffix $n+1$ is identified with $1$ ($x_{n+1}:=x_{1},g_{n+1}:=g_1,y_{n+1}:=y_1$ and $h_{n+1}:=h_1$).
Here $\mathcal{F}_i$ and $\mathcal{F}'_i$ are polynomials of \eqref{fn der x} and \eqref{fn der y}, respectively.
For later convenience, we express $\mathcal{F}_i$ and $\mathcal{F}'_i$ in the standard form:
\begin{align}
\mathcal{F}_i&=c^i_{IJK}T_i^I(\{ x_j\} )S_i^J(\{ g_j\} )D_i^K(\{ \partial /\partial x_j\} ,\{ O^{(g_j)}\} ),\\
\mathcal{F}'_i&=c'^i_{IJK}T_i'^I(\{ y_j\} )S_i'^J(\{ h_j\} )D_i'^K(\{ \partial /\partial y_j\} ,\{ O^{(h_j)}\} ),
\end{align}
where $T_i^I$ is a polynomial of $A_{(I)J}(x_1),\cdots ,A_{(I)J}(x_n)$ and their derivatives, $S_i^J$ is a function of $g_1,\cdots ,g_n$, and $D_i^K$ is a polynomial of $\pder{}{x_1^a},\cdots ,\pder{}{x_n^a},\, O_{ab}^{(g_1)},\cdots ,O_{ab}^{(g_n)}$.
Here the indices $I,J,K$ stand for sets of indices without parentheses.
Hereafter, we do not keep the indices with parentheses, because they are merely labels and have nothing to do with the Lorentz transformations.
$c^i_{IJK}$ is a $Spin(D-1,1)$ invariant tensor because indices without parentheses are contracted to a singlet in \eqref{A0exp}.
We have a similar structure for $\mathcal{F}'_i$.

We choose one coordinate on each index loop.
For simplicity, we choose $x_n$ and $y_n$.
Then we expand the background fields $A_{(I)J}$ around $x_n$ and $y_n$:
\begin{align}
A_{(I)J}(x_i) 
&=\sum_{s=0}^\infty \frac{1}{s!}
\hat A_{(I)J\, a_1\cdots a_s}(x)
\tilde x_i^{a_1}\cdots \tilde x_i^{a_s},
\label{T1}\\
A_{(I)J}(y_i) 
&=\sum_{s=0}^\infty \frac{1}{s!}
\hat A_{(I)J\, a_1\cdots a_s}(y) 
\tilde y_i^{a_1}\cdots \tilde y_i^{a_s},
\label{T2}
\end{align}
where $i=1,\cdots ,n-1$, and 
\begin{align}
\hat A_{(I)J\, a_1\cdots a_s}
=\partial_{a_1}\cdots \partial_{a_s}A_{(I)J}.
\label{hatA}
\end{align}
Here we have introduced the relative coordinate by
\begin{align}
&\tilde x_i=x_i-x_n, \;\;\; \tilde y_i=y_i-y_n \;\;\; \mbox{for} \;\; i=1,\cdots, n-1,
\label{change of variables rel}
\end{align}
and rewritten $x_n$ and $y_n$ as
\begin{align}
&x=x_n,\;\;\; y=y_n.
\label{change of variables n}
\end{align}

Then the propagator in \eqref{n-vertex loop} is expressed in terms of $\tilde x_i$:
\begin{align}
&\mathcal{G}(x_i,g_i;y_i;h_i|x_{i+1},g_{i+1};y_{i+1},h_{i+1}) \nonumber\\
&=G(\tilde x_i-\tilde x_{i+1})
\delta(R^{(a)}_{\;\;\;\;\;b}(g_i^{-1})(\tilde x^b_i-\tilde x^b_{i+1})-R^{(a)}_{\;\;\;\;\;b}(h_i^{-1})(\tilde y^b_i-\tilde y^b_{i+1}))
\delta_{g_ig_{i+1}}\delta_{h_ih_{i+1}},
\label{propa3}
\end{align}
where $i$ runs from 1 to $n$, and $\tilde x_n:=0$, $\tilde x_{n+1}:=\tilde x_1$, $\tilde y_n:=0$ and $\tilde y_{n+1}:=\tilde y_1$.
Note that the right-hand side of (\ref{propa3}) is independent of $x$ and $y$.
This reflects the fact that the translation invariance holds independently for the $x$-system and $y$-system.
Furthermore, the derivatives with respect to $x_i$ and $y_i$ $(i=1,\cdots ,n-1)$ are reduced to those for $\tilde x_i$ and $\tilde y_i$
because we have
\begin{align}
&\frac{\partial}{\partial x^a_i}=\frac{\partial}{\partial \tilde x^a_i},&&
\frac{\partial}{\partial x^a_n}=\frac{\partial}{\partial x^a}-\sum_{i=1}^{n-1}\frac{\partial}{\partial \tilde x^a_i}, &&\mbox{for} \;\; i=1,\cdots, n-1,\nonumber\\
&\frac{\partial}{\partial y^a_i}=\frac{\partial}{\partial \tilde y^a_i},&&
\frac{\partial}{\partial y^a_n}=\frac{\partial}{\partial y^a}-\sum_{i=1}^{n-1}\frac{\partial}{\partial \tilde y^a_i}.&&
\label{change of derivatives}
\end{align}
and $\mathcal{G}$ does not depend on $x$ or $y$.
Therefore we have found that the $x,y$-dependence appears only in the differential polynomials of the background fields.
By separating them from the other part, we found that \eqref{n-vertex loop} is given by a sum of terms as
\begin{align}
&C^{I_1\cdots I_n}_IC'^{J_1\cdots J_n}_J\int \dif{^Dx}\, \dif{^Dy}\, (\hat A(x)\cdots \hat A(x))^I\times (\hat A(y)\cdots \hat A(y))^J\nonumber \\
&\times \int \dif{^D\tilde x_{1}}\cdots \dif{^D\tilde x_{n-1}}\, \dif{^D\tilde y_1}\cdots \dif{^D\tilde y_{n-1}}\,
\dif{g_{1}}\cdots \dif{g_{n}}\, \dif{h_1}\cdots \dif{h_{n}}\, \nonumber \\
&\;\;\;\;\; \times \prod_{i=1}^n \mathcal{\tilde F}_{i\; I_i}(\{ \tilde x_j\} ,\{ g_j\} )\mathcal{\tilde F}'_{i\; J_i}(\{ \tilde y_j\} ,\{ h_j\} )\mathcal{G}(x_{i},g_{i};y_{i},h_{i}|x_{i+1},g_{i+1};y_{i+1},h_{i+1}),
\label{n-vertex loop 2}
\end{align}
where $C^{I_1\cdots I_n}_I$ and $C'^{J_1\cdots J_n}_J$ are $Spin(D-1,1)$ invariant tensors, and $(\hat A(x)\cdots \hat A(x))^I$ and $(\hat A(y)\cdots \hat A(y))^J$ are products of \eqref{hatA} at $x$ and $y$, respectively.
Here $\mathcal{\tilde F}_{i\; I_i}(\{ \tilde x_j\} ,\{ g_j\} )$ and $\mathcal{\tilde F}'_{i\; J_i}(\{ \tilde y_j\} ,\{ h_j\} )$ are
given by
\begin{align}
\mathcal{\tilde F}_{i\; I_i}(\{ \tilde x_j\} ,\{ g_j\} )&=\tilde c^i_{I_iIJK}\tilde f_i^I(\{ \tilde x_j\} )\tilde S_i^J(\{ g_j\} )\tilde D_i^K(\{ \partial /\partial \tilde x_j\} ,\{ O^{(g_j)}\} ),\\
\mathcal{\tilde F}'_{i\; J_i}(\{ \tilde y_j\} ,\{ h_j\} )&=\tilde c'^i_{J_iIJK}\tilde f_i'^I(\{ \tilde y_j\} )\tilde S_i'^J(\{ h_j\} )\tilde D_i'^K(\{ \partial /\partial \tilde y_j\} ,\{ O^{(h_j)}\} ),
\end{align}
where $\tilde f_i^I$ is a polynomial of $\tilde x_1^a,\cdots ,\tilde x_{n-1}^a$, 
$\tilde S_i^J$ is a function of $g_1,\cdots ,g_n$,  and $\tilde D_i^K$ is a polynomial of $\pder{}{\tilde x_1^a},\cdots ,\pder{}{\tilde x_{n-1}^a}$,
$O_{ab}^{(g_1)},\cdots ,O_{ab}^{(g_n)}$.
Here the indices $I,J,K$ stand for sets of indices without parentheses,
and $\tilde c^i_{I_iIJK}$ is a $Spin(D-1,1)$ invariant tensor.
We have a similar structure for $\mathcal{\tilde F}'_{i\; J_i}$.
Note that indices related to $x$ and $g$ and those to $y$ and $h$ form a singlet independently.
This is  because the Taylor expansion and the change of variables do not mix $x$ and $g$ to $y$ and $h$.

In order to examine the integration for the relative coordinates and fiber coordinates,
we consider two independent Lorentz transformations,
\begin{align}
&\tilde x^a_i \rightarrow R^a_{\;\;\;b}(u)\tilde x^b_i \;\;\; \mbox{for} \;\; i=1,\cdots,n-1, \nonumber\\
&g_i \rightarrow ug_i \;\;\; \mbox{for} \;\; i=1,\cdots,n,
\label{Lorentz transformation 1}
\end{align}
and
\begin{align}
&\tilde y^a_i \rightarrow R^a_{\;\;\;b}(v)\tilde y^b_i \;\;\; \mbox{for} \;\; i=1,\cdots,n-1, \nonumber\\
&h_i \rightarrow vh_i \;\;\; \mbox{for} \;\; i=1,\cdots,n,
\label{Lorentz transformation 2}
\end{align}
where $u,\; v\in Spin(D-1,1)$.
The indices in $I_1,\cdots ,I_n$ that appear in \eqref{n-vertex loop 2} come from $\tilde x_1^a,\cdots ,\tilde x_{n-1}^a$, 
$\pder{}{\tilde x_1^a},\cdots ,\pder{}{\tilde x_{n-1}^a}$, $R_{\bb r\kk \beta}^{\;\;\;\;\;\; (\alpha )}(g_1), \cdots ,R_{\bb r\kk \beta}^{\;\;\;\;\;\; (\alpha )}(g_n)$, or $O_{ab}^{(g_1)},\cdots ,O_{ab}^{(g_n)}$.
Under \eqref{Lorentz transformation 1}, each of them transforms  as a corresponding representation.
Similarly, the indices in $J_1,\cdots ,J_n$ transform properly under \eqref{Lorentz transformation 2}.
Under the transformations (\ref{Lorentz transformation 1}) and (\ref{Lorentz transformation 2}),
the integral measure in (\ref{n-vertex loop 2})
and $\mathcal{G}$ is invariant
as is seen from \eqref{propa3}.
Hence, in \eqref{n-vertex loop 2}, the integrations over $\tilde x_1,\cdots ,\tilde x_{n-1}$, $g_1,\cdots ,g_n$, $\tilde y_1,\cdots ,\tilde y_{n-1}$ and
$h_1,\cdots ,h_n$ yield an $Spin(D-1,1)\times Spin(D-1,1)$ invariant tensor.
The indices in $I_1,\cdots ,I_n$ form a singlet and those in $J_1,\cdots ,J_n$ form another singlet.
Then, all the indices in $I$ are contracted to a singlet among themselves, and similarly for $J$.
Thus, (\ref{n-vertex loop 2}) results in the form
\begin{align}
\int \dif{^Dx}\, \dif{^Dy}\; (\hat{A}(x)\cdots \hat{A}(x))\times
(\hat{A}(y)\cdots \hat{A}(y)),
\label{product}
\end{align}
where each factor is a scalar differential polynomial of the background fields.

The effective action is invariant under the $U(N)$ transformation for the background fields which includes
the diffeomorphism and the local Lorentz symmetry.
Hence, by summing the above products of the Lorentz scalars in (\ref{product}),
we obtain, as the total one-loop contribution to the effective
action, 
\begin{align}
\sum_{i,j}c_{ij}s_is_j
\end{align}
where $s_i$'s are diffeomorphism and local Lorentz invariant action functionals, and
$c_{ij}$ are certain constants.
Thus we have proved the factorization at the one-loop level.

For the higher order contributions,
we merely need to consider the cubic and quartic interaction terms in $\phi$.
However the argument proceeds completely parallel to the one-loop case,
and the effective action becomes a product of local actions, each of which comes from each index loop.
Hence, we can show that the factorization occurs to all orders and the effective action takes the form
\begin{align*}
S_{eff}=\sum_i c_{i}s_{i}+\sum_{i,j}c_{ij}s_{i}s_{j}+\sum_{i,j,k}c_{ijk}s_is_js_k+\cdots .
\end{align*}
The degree of each term in $s_i$ is equal to the number of index loops. 
Therefore, the higher order diagrams give the higher-degree polynomials of $s_i$.

\section{Conclusion and discussion}
In this paper, we proved the factorization of the effective action in the derivative interpretation 
of the IIB matrix model.
The point of the proof is that there is the translational invariance on every patch and 
that the tensor indices form singlets on each index loop due to the integral over the fiber coordinates.

Each diagonal block in the classical solution 
is considered to represent a different universe.
The factorization implies that the universes
interact each other by the products of action functionals.
It is also known
that, in the ordinary theory of quantum gravity,
the effective action  on the multiverse takes such a factorized form due to the wormholes,
which are quantum fluctuation of the space-time.
It is expected that the form of the effective action \eqref{S_eff} is universal to the wide class of theories that contain quantum gravity.

Although the action of this type is not local, it does not cause a violation of causality.
Actually, an observer sitting in a universe finds that the nature is described by an ordinary local field theory.
The reason is as follows.
The path-integral based on the action \eqref{S_eff} can be rewritten as 
\begin{align}
Z&=\int \mathcal{D}\phi \, e^{i(\sum_i c_is_i+\cdots )}\nonumber \\
&=\int \dif{\lambda}\, f(\lambda )\int \mathcal{D}\phi\, e^{i\sum_i \lambda _is_i},
\end{align}
where $f(\lambda )$ is a function of the coupling constants $\lambda _1,\lambda _2,\cdots$.
Therefore, if $\lambda _i$'s are fixed to some values by some mechanism, the observer in a universe merely sees a local theory given by
\begin{align}
S=\sum_i \lambda _is_i.
\end{align}
Although this type of action has been discussed from various points of view \cite{S82Ka,L88Un,C88Wh,Yoneya:2008ka}, it seems that there is no definite understanding so far.
In particular, it is interesting to see whether the values of $\lambda _i$'s depend on the initial conditions or are fixed dynamically to some special values.
If it is the latter case, we may construct a model which describes the real world with realistic parameters.
For example, referring to the works \cite{CSZ11In,Aoki:2010gv}, 
we can show that
the particle content in the standard model is realized around classical solutions in the derivative interpretation of the IIB matrix model,
and it would be interesting if all the coupling constants are fixed by such mechanism.

\section*{Acknowledgment}
This work is supported by the Grant-in-Aid for the Global COE Program The Next Generation of Physics, Spun from Universality and Emergence from the Ministry of Education, Culture, Sports, Science and Technology (MEXT) of Japan.
The work of A.T. is supported in part
by Grant-in-Aid for Scientific Research 
(Nos. 24540264 and 23244057) from JSPS.

\appendix

\section{Example of factorization}
In this appendix, we give an example of the calculation of the effective action at the one-loop level.

\begin{figure}[htbp]
\begin{center}
\includegraphics[scale=0.33]{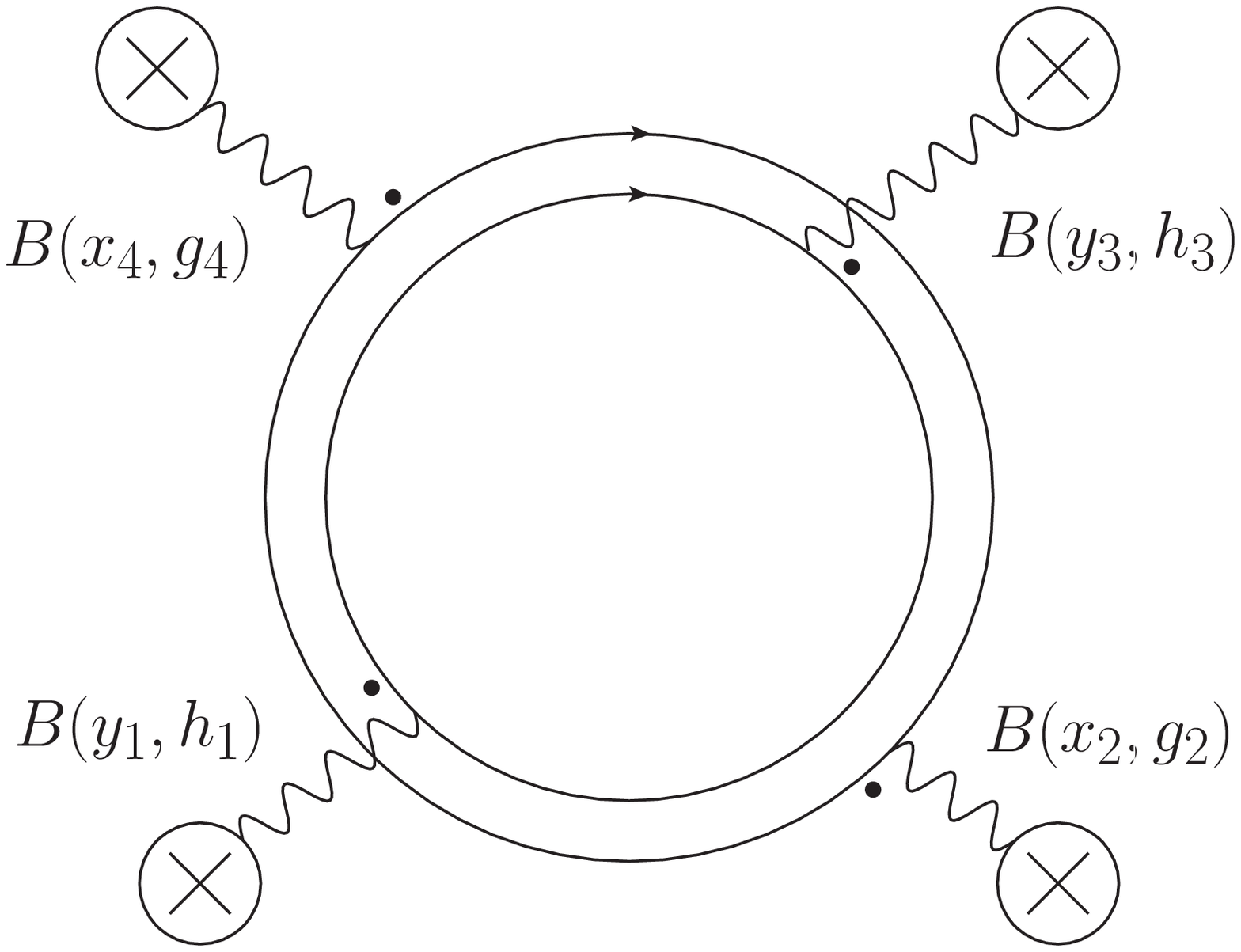}
\caption{\small A four-point diagram for $B_{a}$. 
$\bullet$ indicates the side on which the derivative $(\pder{}{\xi}+\pder{}{\eta})$ in the kinetic part acts.}
\label{fig:B}
\end{center}
\end{figure}

Here we consider a four-point function of $B_{a}(x)$
depicted in Figure $\ref{fig:B}$.
This diagram is calculated as follows:
\begin{align}
&-i\int \dif{^Dx _1}\cdots \dif{^Dx _4}\,  \dif{^Dy _1}\cdots \dif{^Dy _4}\, \dif{g_1}\cdots \dif{g_4}\, \dif{h_1}\cdots \dif{h_4}\nonumber \\
&\; \times B^{(a''')}(y_3,h_3)(R^{a}_{\;\; (a)}(g_4)\pder{}{x_4^{a}}+R^{a}_{\;\; (a)}(h_4)\pder{}{y_4^{a}})\mathcal{G}(x_3,g_3;y_3,h_3|x_4,g_4;y_4,h_4)\nonumber \\
&\;\; \times B^{(a)}(x_4,g_4)(R^{a'}_{\;\; (a')}(g_1)\pder{}{x_1^{a'}}+R^{a'}_{\;\; (a')}(h_1)\pder{}{y_1^{a'}})\mathcal{G}(x_4,g_4;y_4,h_4|x_1,g_1;y_1,h_1)\nonumber \\
&\;\;\; \times B^{(a')}(y_1,h_1)(R^{a''}_{\;\; (a'')}(g_2)\pder{}{x_2^{a''}}+R^{a''}_{\;\; (a'')}(h_2)\pder{}{y_2^{a''}})\mathcal{G}(x_1,g_1;y_1,h_1|x_2,g_2;y_2,h_2)\nonumber \\
&\;\;\;\; \times B^{(a'')}(x_2,g_2)(R^{a'''}_{\;\; (a''')}(g_3)\pder{}{x_3^{a'''}}+R^{a'''}_{\;\; (a''')}(h_3)\pder{}{y_3^{a'''}})\mathcal{G}(x_2,g_2;y_2,h_2|x_3,g_3;y_3,h_3),
\label{4ptB}
\end{align}
where $\mathcal{G}(x_i,g_i;y_i,h_i|x_{i+1},g_{i+1};y_{i+1},h_{i+1})$ is given by \eqref{propa2}.
The action of $R^{a}_{\; (a)}(g_i)\pder{}{x_i^{a}}+R^{a}_{\; (a)}(h_i)\pder{}{y_i^{a}}$ on $\mathcal{G}$ can be evaluated as
\begin{align}
&\left( R^{a}_{\; (a)}(g_i)\pder{}{x_i^{a}}+R^{a}_{\; (a)}(h_i)\pder{}{y_i^{a}}\right) \mathcal{G}(x_i,g_i;y_i,h_i|x_{i+1},g_{i+1};y_{i+1},h_{i+1}) \nonumber \\
&\;\; =-R^{a}_{\;\; (a)}(g_i)\partial _{a}G(x_{i}-x_{i+1}) \delta (R^{(b)}_{\;\;\;\, c}(g_{i}^{-1})(x_{i}^c-x_{i+1}^c)-R^{(b)}_{\;\;\;\, c}(h_{i}^{-1})(y_{i}^c-y_{i+1}^c))\delta _{g_ig_{i+1}}\delta _{h_ih_{i+1}}.
\label{delG}
\end{align}

As in \eqref{change of variables rel} and \eqref{change of variables n}, we change the variables for $n=4$, and obtain the chain rules \eqref{change of derivatives}.
The integrations over the fiber coordinates $g_1,g_2$ and $g_3$ and $h_1,h_2$ and $h_3$ are trivially performed because there is no $O_{bc}$ inserted,
and we have a constant factor $\gamma ^2$.
Here $\gamma$ is a divergent constant that represents the value of the delta function when the two points coincide: $\gamma =\delta _{gg}$.
Then, making the Taylor expansion, we obtain
\begin{align}
&\sum_{s,m,u=0}^\infty \frac{\gamma ^2}{s!t!u!}\int \dif{^Dx}\, B^{a}(x)\partial _{d_1}\cdots \partial _{d_u}B^{a''}(x) \int \dif{^Dy}\, \partial _{b_1}\cdots \partial _{b_s}B^{a'''}(y)\partial _{c_1}\cdots \partial _{c_t}B^{a'}(y)\nonumber \\
&\;\; \times \int \dif{g}\, R_{\;\; (a')}^{e'}(g)R_{\;\; (a''')}^{e'''}(g) \int \dif{h}\, R^{(a')}_{\;\;\;\;\; a'}(h^{-1})R^{(a''')}_{\;\;\;\;\; a'''}(h^{-1})\nonumber \\
&\;\;\;\;\; \times (-i)\int \dif{^D\tilde x_1}\cdots \dif{^D\tilde x_3}\; \dif{^D\tilde y_1}\cdots \dif{^D\tilde y_3}\; \tilde y_3^{b_1}\cdots \tilde y_3^{b_s}\tilde y_1^{c_1}\cdots \tilde y_1^{c_t}\tilde x_2^{d_1}\cdots \tilde x_2^{d_u}\nonumber \\
&\;\;\;\;\;\;\;\;\;\;\;\;\;\;\;\;\; \times \partial _{a}G(\tilde x_3)\partial _{e'}G(-\tilde x_1) \partial _{a''}G(\tilde x_1-\tilde x_2)\partial _{e'''}G(\tilde x_2-\tilde x_3)\nonumber \\
&\;\;\;\;\;\;\;\;\;\;\;\;\;\;\;\;\; \times \delta (\tilde \xi _3-\tilde \eta _3)\delta (\tilde \xi_1-\tilde \eta _1) \delta ((\tilde \xi _1-\tilde \xi _2)-(\tilde \eta _1-\tilde \eta _2))\delta ((\tilde \xi _2-\tilde \xi _3)-(\tilde \eta _2-\tilde \eta _3)),
\label{4ptB2}
\end{align}
where $g:=g_4$, $h:=h_4$, and $\tilde \xi _i$ and $\tilde \eta _i$ are defined by
\begin{align}
&\tilde \xi _i^{(a)}:=R^{(a)}_{\;\;\;\;\; b}(g^{-1})\tilde x_i^b\; ,&&\tilde \eta _i^{(a)}:=R^{(a)}_{\;\;\;\;\; b}(h^{-1})\tilde y_i^b.
\end{align}
$\tilde y_i$-dependence in the integrand appears only in the product of the delta functions. Since one of the four delta functions in \eqref{4ptB2} is dependent on the others, it is proportional to $\delta (0)$.

Since the product of the second to the fifth line in \eqref{4ptB2} is invariant under two independent Lorentz transformations (\eqref{Lorentz transformation 1} and \eqref{Lorentz transformation 2}), it forms a $Spin(D-1,1)\times Spin(D-1,1)$ invariant tensor.
Because it is proportional to $\delta (0)$, as we have seen, it can be written as
\begin{align}
\delta (0)C^{d_1\cdots d_u}_{aa''}C'^{b_1\cdots b_s c_1\cdots c_t}_{a'a'''}.
\end{align}
Therefore \eqref{4ptB2} turns out to be
\begin{align}
&\sum_{s,m,u=0}^\infty \frac{\gamma ^2\delta (0)}{s!t!u!}\int \dif{^Dx}\, C^{d_1\cdots d_u}_{aa''}B^{a}(x)\partial _{d_1}\cdots \partial _{d_u}B^{a''}(x)\nonumber \\
&\;\;\;\;\;\;\;\;\;\;\;\; \times \int \dif{^Dy}\, C'^{b_1\cdots b_s c_1\cdots c_t}_{a'a'''}\partial _{b_1}\cdots \partial _{b_s}B^{a'''}(y)\partial _{c_1}\cdots \partial _{c_t}B^{a'}(y),
\end{align}
which is apparently factorized.

\section{Spin-one and spin-two fields}
In order for the model to be realistic, there must be massless
spin-one and spin-two fields (gauge fields and gravitons).
We assume that the higher spin fields are massive, so that we do not have to consider the mixing of spin-one or spin-two fields to them.

\begin{figure}[htbp]
\begin{center}
\includegraphics[scale=0.55]{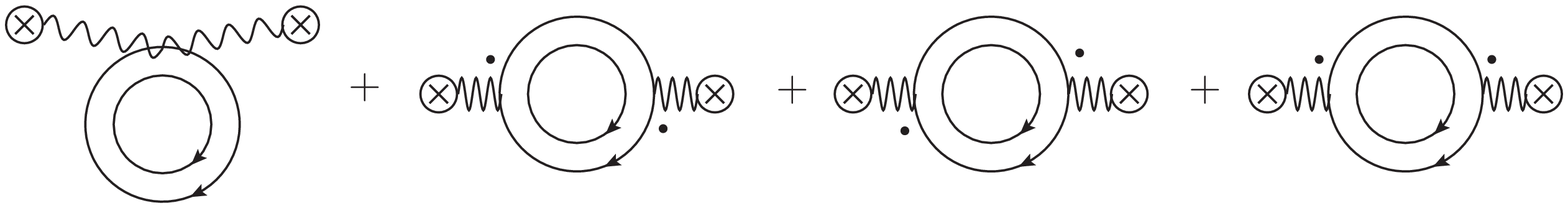}
\caption{\small Diagrams for the mass term of $B_{a}(x)$.}
\label{fig:Bmass}
\end{center}
\end{figure}

First, we consider the spin-one vector $B_{a}(x)$.
The mass term of $B_{a}(x)$ in the one-loop effective action is depicted in Figure \ref{fig:Bmass}, where
we have two insertions of $B_{a}(x)$'s in the index loop specified by $(x,g)$ and 
no insertions in the other index loop specified by $(y,h)$.
The sum of the diagrams is given by
\begin{align}
&-i\sum_{n=0}^\infty \frac{\delta (0)\gamma ^2V_{\mathcal{M}}V_{G}}{n!}\int \dif{^Dx}\, \dif{g}\, B^a(x)\partial _{b_1}\cdots \partial _{b_n}B^{a'}(x)\int \dif{^D\tilde x_1}\, \tilde x_1^{b_1}\cdots \tilde x_1^{b_n}\nonumber \\
&\;\;\;\;\;\; \times \Big( \frac{1}{2}\partial _{a}G(\tilde x_1)\partial _{a'}G(-\tilde x_1)+\frac{1}{2}\partial _{a'}G(\tilde x_1)\partial _{a}G(-\tilde x_1)-G(\tilde x_1)\partial _{a}\partial _{a'}G(-\tilde x_1) \Big) \nonumber \\
&\; -i\delta (0)\gamma ^2V_{\mathcal{M}}V_{G}^2\int \dif{^Dx}\, B^a(x)B_{a}(x)G(0),
\label{2ptB}
\end{align}
where $V_{\mathcal{M}}$ is the volume of $\mathcal{M}$,
and $V_G$ is the volume of the fiber $Spin(D-1,1)$.
Since the integrations over $x$ and $y$ are completely decoupled, 
(\ref{2ptB}) is reduced to the result in the ordinary scalar quantum electrodynamics.
While the $n=2$ term in the Taylor series gives the one-loop corrections to the Maxwell term, the $n=0$ term is the mass term, which vanishes due to the gauge invariance.

In a similar way, we can calculate the two-point function of the spin-two gravitational field $h_{a}^{\;\;\; b}(x)$.
The result for $h_{a}^{\;\;\; b}(x)$ is also reduced to that in the ordinary quantum field theory.
However, this is a little nontrivial
because $h_{a}^{\;\;\; b}(x)$ associates a derivative, which acts also on the delta functions $\delta (\tilde \xi ' -\tilde \eta ')$ in the propagators.
Explicitly, the lowest order of the Taylor series of the quadratic term for $h_a^{\;\;\; b}(x)$ can be calculated as
\begin{align}
&i\delta (0)\gamma ^2V_{\mathcal{M}}V_{G}\int \dif{^Dx}\, \dif{g}\, h^{ab}(x)h^{a'b'}(x)\Big\{ i\eta _{aa'}\pder{}{x^{b}}\pder{}{x'^{b'}}\left[ G(x-x')\delta (x-x')\right] \Big| _{x'\to x}\nonumber \\
&\;\;+\int \dif{^D\tilde x_1}\, \dif{^D\tilde y_1}\, \Big( \frac{1}{2}\pder{}{\tilde x_1^{b'}}\Big[ (\pder{}{\tilde x_1^{a}}G(\tilde x_1))\delta (\tilde x_1-\tilde y_1)\Big] \pder{}{\tilde x_1^{b}}\Big[ (\pder{}{\tilde x_1^{a'}}G(-\tilde x_1))\delta (\tilde x_1-\tilde y_1)\Big] \nonumber \\
&\;\;\;\;\; +\frac{1}{2}\pder{}{\tilde x_1^{b}}\Big[ (\pder{}{\tilde x_1^{a'}}G(\tilde x_1))\delta (\tilde x_1-\tilde y_1)\Big] \pder{}{\tilde x_1^{b'}}\Big[ (\pder{}{\tilde x_1^{a}}G(-\tilde x_1))\delta (\tilde x_1-\tilde y_1)\Big] \nonumber \\
&\;\;\;\;\;\;\;\; +\pder{}{\tilde x_1^{b}}\pder{}{\tilde x_1^{b'}}\Big[ G(\tilde x_1)\delta (\tilde x_1-\tilde y_1)\Big] \; (\partial _{a}\partial _{a'}G(-\tilde x_1)) \delta (\tilde x_1-\tilde y_1)\Big) \Big\} .
\label{4ptGr}
\end{align}
The Leibniz rule yields three classes of terms.
In the first class, two derivatives act on the delta functions. These terms do not correspond to the ordinary field theory of $h_{a}^{\;\;\; b}(x)$.
However, we can show that they are combined to zero if we use the same identity as we have used to show the vanishing of the vector mass above.
In the second class, one derivative acts on $G$ and the other on the delta function.
These terms vanish because of the $\tilde y_1$-integration.
In the third class, two derivatives act on $G$.
These terms are the same as the ordinary field theory.
They are combined to the form
\begin{align}
\int \dif{^Dx}\left\{ \left( h^a_a(x)\right) ^2+h^{ab}(x)h_{ba}(x)\right\} ,
\end{align}
which is the quadratic part of the cosmological constant.
Thus we have seen that \eqref{4ptGr} gives the part of 
$c\int \dif{^Dx}\sqrt{-g(x)}\int \dif{^Dy}\sqrt{-g(y)}$ which is the zeroth order in $h_{a}^{\;\;\; b}(y)$ and the second order in $h_{a}^{\;\;\; b}(x)$.


\end{document}